\documentclass{emulateapj}
\usepackage{apjfonts}
\usepackage[]{natbib}
\usepackage{graphics}

\newcommand{\etal}{et al.}  
\newcommand{\per}{\ensuremath{^{-1}}}
\newcommand{\persq}{\ensuremath{^{-2}}}

\newcommand{\hbeta}{H\ensuremath{\beta}} 
\newcommand{\hgamma}{H\ensuremath{\gamma}} 
 
\newcommand{\heii}{\ion{He}{2}}

\newcommand{\msun}{\ensuremath{{M}_{\odot}}}
 
\newcommand{\kms}{km~s\ensuremath{^{-1}}}

\newcommand{\mbh}{\ensuremath{M_\mathrm{BH}}}

\newcommand{\sigmastar}{\ensuremath{\sigma_\star}}

\newcommand{\msigma}{\ensuremath{\mbh-\sigma}}
\newcommand{\rblr}{\ensuremath{r_{\mathrm{BLR}}}}

\newcommand{\tpeak}{\ensuremath{\tau_\mathrm{peak}}}
\newcommand{\tcen}{\ensuremath{\tau_\mathrm{cen}}}

\shorttitle{REVERBERATION MAPPING OF MRK 50}
\shortauthors{BARTH ET AL.}

\begin{document} 

\title{The Lick AGN Monitoring Project 2011:  Reverberation Mapping
of Markarian 50}

\author{
  Aaron J. Barth\altaffilmark{1}, 
  Anna Pancoast\altaffilmark{2}, 
  Shawn J. Thorman\altaffilmark{1}, 
  Vardha N. Bennert\altaffilmark{2,3}, 
  David J. Sand\altaffilmark{2,4}, 
  Weidong Li\altaffilmark{5}, 
  Gabriela Canalizo\altaffilmark{6}, 
  Alexei V. Filippenko\altaffilmark{5},
  Elinor L. Gates\altaffilmark{7}, 
  Jenny E. Greene\altaffilmark{8}, 
  Matthew A. Malkan\altaffilmark{9},
  Daniel Stern\altaffilmark{10}, 
  Tommaso Treu\altaffilmark{2}, 
  Jong-Hak Woo\altaffilmark{11},
  Roberto J. Assef\altaffilmark{10,12}, 
  Hyun-Jin Bae\altaffilmark{13}, 
  Brendon J. Brewer\altaffilmark{2},
  Tabitha Buehler\altaffilmark{14}, 
  S. Bradley Cenko\altaffilmark{5}, 
  Kelsey I. Clubb\altaffilmark{5}, 
  Michael C. Cooper\altaffilmark{1,15}, 
  Aleksandar M. Diamond-Stanic\altaffilmark{16,17}, 
  Kyle D. Hiner\altaffilmark{6}, 
  Sebastian F. H\"{o}nig\altaffilmark{2},
  Michael D. Joner\altaffilmark{14}, 
  Michael T. Kandrashoff\altaffilmark{5}, 
  C. David Laney\altaffilmark{14}, 
  Mariana S. Lazarova\altaffilmark{6}, 
  A. M. Nierenberg\altaffilmark{2},
  Dawoo Park\altaffilmark{11}, 
  Jeffrey M. Silverman\altaffilmark{5,18},
  Donghoon Son\altaffilmark{11}, 
  Alessandro Sonnenfeld\altaffilmark{2},
  Erik J. Tollerud\altaffilmark{1},
  Jonelle L. Walsh\altaffilmark{1,19},
  Richard Walters\altaffilmark{20}, 
  Robert L. da Silva\altaffilmark{21}, 
  Michele Fumagalli\altaffilmark{21}, 
  Michael D. Gregg\altaffilmark{22}, 
  Chelsea E. Harris\altaffilmark{2},
  Eric Y. Hsiao\altaffilmark{18}, 
  Jeffrey Lee\altaffilmark{17}, 
  Liliana Lopez\altaffilmark{17}, 
  Jacob Rex\altaffilmark{5}, 
  Nao Suzuki\altaffilmark{18}, 
  Jonathan R. Trump\altaffilmark{21},
  David Tytler\altaffilmark{17}, 
  G\'{a}bor Worseck\altaffilmark{21}, 
  and
  Hassen M. Yesuf\altaffilmark{21}
}

\altaffiltext{1}{Department of Physics and Astronomy, 4129 Frederick
  Reines Hall, University of California, Irvine, CA, 92697-4575, USA;
  barth@uci.edu}

\altaffiltext{2}{Department of Physics, University of California,
Santa Barbara, CA 93106, USA}

\altaffiltext{3}{Physics Department, California Polytechnic State
University, San Luis Obispo, CA 93407, USA}

\altaffiltext{4}{Las Cumbres Observatory Global Telescope Network,
  6740 Cortona Drive, Suite 102, Santa Barbara, CA 93117, USA}

\altaffiltext{5}{Department of Astronomy, University of California,
Berkeley, CA 94720-3411, USA}

\altaffiltext{6}{Department of Physics and Astronomy, University of
  California, Riverside, CA 92521, USA}

\altaffiltext{7}{Lick Observatory, P. O. Box 85, Mount Hamilton, CA
95140, USA}

\altaffiltext{8}{Department of Astrophysical Sciences, Princeton
  University, Princeton, NJ 08544, USA}

\altaffiltext{9}{Department of Physics and Astronomy, University of
California, Los Angeles, CA 90095-1547, USA}

\altaffiltext{10}{Jet Propulsion Laboratory, California Institute of
Technology, 4800 Oak Grove Boulevard, Pasadena, CA 91109, USA}

\altaffiltext{11}{Astronomy Program, Department of Physics and
Astronomy, Seoul National University, Seoul 151-742, Republic of
Korea}

\altaffiltext{12}{NASA Postdoctoral Program Fellow}

\altaffiltext{13}{Department of Astronomy and Center for Galaxy
Evolution Research, Yonsei University, Seoul 120-749, Republic of
Korea}

\altaffiltext{14}{Department of Physics and Astronomy, N283 ESC,
Brigham Young University, Provo, UT 84602-4360, USA}

\altaffiltext{15}{Hubble Fellow}

\altaffiltext{16}{Southern California Center for Galaxy Evolution Fellow}

\altaffiltext{17}{Center for Astrophysics and Space Sciences,
  University of California, San Diego, CA 92093-0424, USA}

\altaffiltext{18}{Physics Division, Lawrence Berkeley National
Laboratory, 1 Cyclotron Road, Berkeley, CA 94720, USA}

\altaffiltext{19}{Department of Astronomy, The University of Texas at
  Austin, Austin, TX 78712, USA}

\altaffiltext{20}{Caltech Optical Observatories, California Institute
of Technology, Pasadena, CA 91125, USA}

\altaffiltext{21}{Department of Astronomy and Astrophysics, UCO/Lick
 Observatory, University of California, 1156 High Street, Santa Cruz,
 CA 95064, USA}

\altaffiltext{22}{Department of Physics, University of California
Davis, Davis, CA 95616, USA; IGPP, Lawrence Livermore National
Laboratory, Livermore, CA 94550, USA}

\begin{abstract}
The Lick AGN Monitoring Project 2011 observing campaign was carried
out over the course of 11 weeks in Spring 2011.  Here we present the
first results from this program, a measurement of the broad-line
reverberation lag in the Seyfert 1 galaxy Mrk 50.  Combining our data
with supplemental observations obtained prior to the start of the main
observing campaign, our dataset covers a total duration of 4.5 months.
During this time, Mrk 50 was highly variable, exhibiting a maximum
variability amplitude of a factor of $\sim4$ in the $U$-band continuum
and a factor of $\sim2$ in the \hbeta\ line.  Using standard
cross-correlation techniques, we find that \hbeta\ and \hgamma\ lag
the $V$-band continuum by $\tcen = 10.64_{-0.93}^{+0.82}$ and
$8.43_{-1.28}^{+1.30}$ days, respectively, while the lag of \heii\
$\lambda4686$ is unresolved.  The \hbeta\ line exhibits a symmetric
velocity-resolved reverberation signature with shorter lags in the
high-velocity wings than in the line core, consistent with an origin
in a broad-line region dominated by orbital motion rather than infall
or outflow.  Assuming a virial normalization factor of $f=5.25$, the
virial estimate of the black hole mass is $(3.2\pm0.5)\times10^7$
\msun.  These observations demonstrate that Mrk 50 is among the most
promising nearby active galaxies for detailed investigations of
broad-line region structure and dynamics.

\end{abstract}


\slugcomment{Accepted for publication in ApJ Letters}

\section{Introduction}

Observations of broad-line variability in nearby Seyfert galaxies play
a central role in the interpretation of the demographics and
cosmological evolution of supermassive black holes in active galactic
nuclei (AGNs).  By measuring the time delay between AGN continuum
variations and the subsequent response of the broad-line region (BLR)
gas, the light-travel time across the BLR, and hence the mean BLR
radius (\rblr), can be directly measured.  These reverberation-mapping
measurements have been carried out for a few dozen low-redshift AGNs
\citep[e.g.,][]{kaspi2000, peterson2004, bentz2009a}, and the observed
BLR sizes measured via \hbeta\ reverberation range from typically a
few light-days up to several light-months.  The BLR size is observed
to scale with AGN continuum luminosity roughly as $\rblr \propto
L^{0.5}$ \citep{bentz2009b}, and this relationship makes it possible
to estimate BLR radii from a single observation of an AGN.

With a direct measurement or estimate of \rblr, and assuming virial
motion of BLR clouds, it becomes possible to estimate the mass of the
black hole in an AGN as $\mbh = f \rblr (\Delta V)^2 / G$, where
$\Delta V$ is the width of the broad line, and $f$ is a dimensionless
scaling factor \citep[e.g.,][]{ulrich1984,kaspi2000,onken2004}.  This
method has been used to estimate black hole masses in large samples of
AGNs out to the highest observed redshifts \citep[for a review,
see][]{vestergaard2011}.  Currently, nearly all observational
constraints on the cosmological growth history of supermassive black
holes depend on masses derived from this virial equation.  While the
assumption of virial motion in the BLR has gained support from a
variety of consistency checks \citep[e.g.,][]{petersonwandel2000,
onken2004, nelson2004}, it remains extremely difficult to obtain
direct constraints on the structure and dynamical state of the BLR in
any individual AGN, and in the absence of such constraints, the
inferred black hole masses remain subject to substantial systematic
uncertainty \citep{krolik2001}.  The most promising method to examine
the kinematics of BLR gas is velocity-resolved reverberation mapping,
in which the time-delay response of emission-line variability relative
to continuum fluctuations can be measured as a function of
line-of-sight velocity.  Velocity-resolved reverberation data can
encode a wealth of information about BLR structure on spatial scales
that are orders of magnitude too small to be resolved by any other
technique \citep[e.g.,][]{welsh1991,horne2004}.

Recently, high-cadence observing campaigns have produced major
improvements in the quality of velocity-resolved reverberation data
for the broad Balmer lines \citep{bentz2009a, denney2010}.  For the
most highly variable objects, it is possible to examine the shape of
the two-dimensional transfer function, which describes the
distribution of broad-line lag response time as a function of velocity
\citep{bentz2010}, and to apply new modeling techniques that can
directly constrain the BLR geometry and potentially test the critical
assumption of virial motion \citep{pancoast2011a,brewer2011}.  In order
to increase the number of objects with data suitable for such
analysis, we carried out a new reverberation-mapping campaign in
Spring 2011.

In this \emph{Letter}, we present our first results for Mrk 50, a
Seyfert 1 galaxy at redshift $z=0.023$.  Past observations have found
dramatic variations in its nuclear luminosity \citep{pastoriza1991},
but it has not previously been a reverberation-mapping target.  During
our monitoring program, Mrk 50 exhibited strong variability, and we
detect a robust reverberation signal and a significant
velocity-resolved reverberation response across the broad \hbeta\
emission line.

\section{Observations and Reductions}

\subsection{Photometry}

From 2011 January 21 until June 13 (all dates are UT), we obtained
queue-scheduled $V$-band images using the 0.76~m Katzman Automatic
Imaging Telescope at Lick Observatory \citep{filippenko2001}, the
0.9~m telescope at the Brigham Young West Mountain Observatory (WMO),
the Super-LOTIS 0.6~m telescope at Kitt Peak, the Faulkes Telescope
South at Siding Spring Observatory, and the Palomar 1.5~m telescope
\citep{cenko2006}.  Exposure times were typically 180--300 s.  We
attempted to observe Mrk 50 on a nightly basis, but poor weather and
telescope scheduling issues left some gaps in temporal coverage.

All images were bias-subtracted and flattened, and cosmic-ray hits
were removed using the LA-COSMIC routine \citep{vandokkum2001}.  In
order to remove the host galaxy and obtain a light curve of the
variable AGN flux, we employed image subtraction using the
\texttt{HOTPANTS} package by
A. Becker\footnote{http://www.astro.washington.edu/users/becker/c\_software.html},
which is based on the algorithm described by \citet{alard1999}.  For
each telescope, a high-quality template image was chosen, and the
template was then aligned with each night's image and convolved with a
spatially varying kernel to match the point-spread function of that
image.  After subtracting the scaled template image, the variable AGN
flux is left as a point source in the subtracted image, allowing for
aperture photometry using the IRAF DAOPHOT package. The photometric
aperture radius used for each telescope was set to match the average
point-source full width at half-maximum intensity (FWHM) for images
from that telescope.  Light curves were constructed separately for
each telescope, and each was then normalized to the WMO light curve by
determining an average flux scaling factor based on nights when Mrk 50
was observed at both locations.  We find that the image subtraction
yields a better-quality light curve than simple aperture photometry,
and provides closer agreement between the light curves obtained with
different telescopes.

\begin{figure}[t!]
\plotone{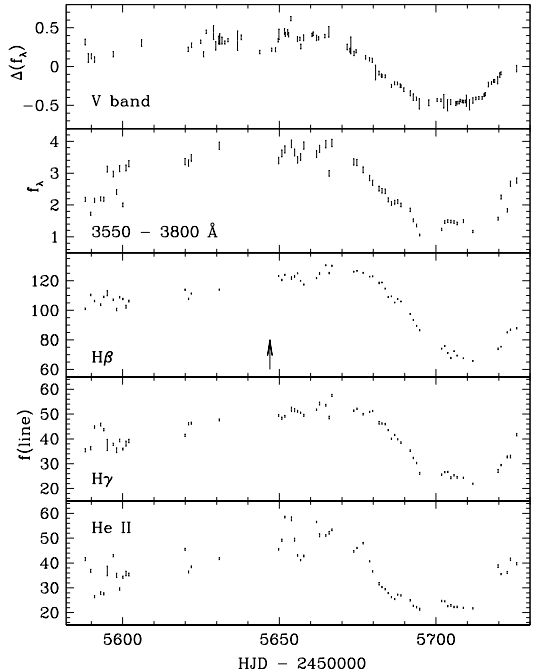}
\caption{Mrk 50 light curves for the $V$ band, the $U_s$-band
  continuum measured from the spectra over 3550--3800 \AA, and the
  \hbeta, \hgamma, and \heii\ emission lines.  The $V$-band panel
  shows a difference imaging light curve illustrating changes in flux
  relative to the mean.  Units for the $V$ and $U_s$ bands are
  $10^{-15}$ erg cm\persq\ s\per\ \AA\per, and units for broad-line
  fluxes are $10^{-15}$ erg cm\persq\ s\per. The arrow in the \hbeta\
  light curve marks the start of the dedicated Lick observing
  campaign.
\label{figlightcurves}}
\end{figure}

The overall flux scale was calibrated using observations of Landolt
(1992) standard stars observed during a few photometric nights
at WMO. Observations taken within 6~hr of one another were combined
by taking a weighted average.  The final $V$ light curve is
illustrated in Figure \ref{figlightcurves}.

\subsection{Spectroscopy}

Our campaign at the Lick Shane 3 m telescope consisted of 69 nights
allocated between 2011 March 27 and June 13, during which time we
observed Mrk 50 on 41 nights using the Kast dual spectrograph
\citep{miller93}.  In this paper, we discuss only measurements from
the blue arm of the spectrograph, where we used a 600 lines mm\per\
grism over $\sim 3440$--5520~\AA\ at a scale of 1.0 \AA\ pixel\per.
All observations were done with a 4\arcsec-wide slit oriented at
$PA=180$\arcdeg.  Standard calibration frames including arc lamps and
dome flats were observed each afternoon, and flux standards were
observed during twilight.  The exposure time for Mrk 50 was normally
$2\times1200$ s. Additionally, we observed Mrk 50 on 18 nights that
were allocated to other observing programs, beginning on January 26.
All observations used the same 600-line grism and 4\arcsec\ slit, but
the wavelength coverage was slightly different each time.  The
exposure for these observations was typically 900~s.

Spectroscopic reductions and calibrations followed standard methods
implemented in IRAF and IDL.  A large extraction width of 10\farcs3
was used in order to accommodate the full extent of the AGN spatial
profiles observed on nights with very poor seeing.  Error spectra were
extracted and propagated through the full sequence of calibrations,
and for each night the two exposures of Mrk 50 were averaged
together. In the region 4600--4800~\AA, the median signal-to-noise
ratio per pixel is 75.

\section{Spectroscopic Data Analysis}

The reduced spectra were first normalized to a uniform flux scale by
employing the procedure of \citet{vgw1992}.  This method applies a
flux scaling factor, a linear wavelength shift, and a Gaussian
convolution to each spectrum in order to minimize the residuals
between the data and a reference spectrum constructed from several of
the best-quality nights.  The scaling is determined using a wavelength
range containing the [\ion{O}{3}] $\lambda5007$ line, which is assumed
to have constant flux.

\begin{figure}[t!]
\plotone{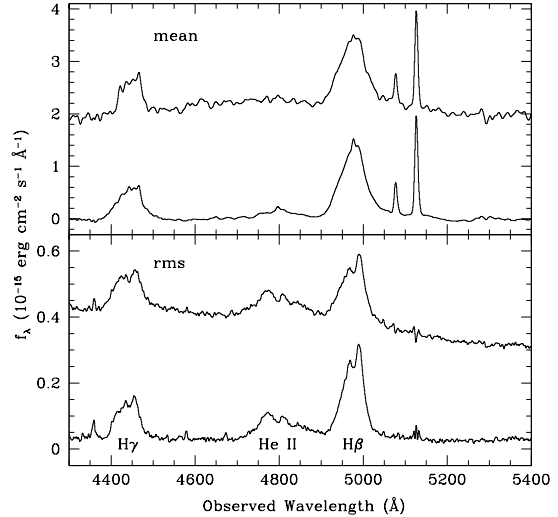}
\caption{Mean and rms spectra.  In each panel, the upper spectrum is
  constructed from the set of scaled spectra of Mrk 50, and the lower
  spectrum is constructed from the continuum-subtracted, scaled
  spectra. 
\label{figmean}}
\end{figure}

Since the spectra contain a substantial amount of host-galaxy
starlight, we implemented a simple continuum subtraction routine to
produce a cleaner measurement of the broad-line profiles and fluxes,
similar to the procedure described by \citet{park2011}.  Over the
wavelength range 4300--5400 \AA, each scaled spectrum was fitted with
a model consisting of a power-law featureless continuum, an
\ion{Fe}{2} template from \citet{borosongreen1992} broadened by
convolution with a Gaussian kernel, several emission-line components
represented by either Gaussians or Gauss-Hermite functions
\citep{vandermarel1993}, and a starlight model consisting of simple
stellar population models at solar metallicity from
\citet{bruzualcharlot2003} which were broadened by convolution with a
Gaussian kernel.  Additionally, a foreground extinction was applied to
the model spectrum.  The model fit was optimized by $\chi^2$
minimization using a Levenberg-Marquardt technique
\citep{markwardt2009}.  Then, the best-fitting model components
representing the starlight and nonstellar continuum were subtracted
from the data, leaving a pure emission-line spectrum. For the
starlight model, we obtained a good fit using an 11 Gyr-old population
as the dominant component, and adding contributions from younger
populations did not significantly improve the fit. The median
starlight fraction at $\lambda_\mathrm{rest}=5100$ \AA\ is 41\%, and
\ion{Fe}{2} contributes just $\sim2$--3\% of the continuum flux
density at 4600~\AA.

In prior reverberation work, starlight subtraction has not generally
been applied, and the broad-line light curves have typically been
measured by subtracting a local, linear continuum fitted to wavelength
regions on either side of an emission line.  We find that our
continuum subtraction procedure provides a more accurate removal of
the continuum shape underlying each emission line and a much better
light curve for the weak \heii\ $\lambda4686$ line.  Figure
\ref{figmean} shows two versions of the mean and root-mean-square
(rms) spectra: one constructed from the set of scaled spectra, and one
constructed from the set of continuum-subtracted, scaled spectra.
This illustrates the utility of the continuum subtraction procedure
for removing the stellar absorption features and continuum
undulations.

The broad-line light curves were then measured by integrating the flux
over fixed wavelength ranges in the continuum-subtracted spectra:
4890--5050 \AA\ for \hbeta, 4370--4510 \AA\ for \hgamma\ (also
including the narrow [\ion{O}{3}] $\lambda4363$ line), and 4730--4870
\AA\ for \heii.  Emission-line light curves are shown in Figure
\ref{figlightcurves}, along with the continuum flux density measured
from the scaled spectra over 3550--3800 \AA\ (which we refer to as the
$U_s$ band).  The continuum and broad lines were highly variable
during the monitoring period: over a 30-day span the $U_s$-band
continuum dropped by a factor of $\sim4$, and the \hbeta\ line
responded with a factor of $\sim2$ decline.  For observations taken
before the start of our dedicated campaign on March 27, the scatter in
the spectroscopic light curves is relatively high.  We attribute this
to the nightly differences in observational setup and calibrations
used by each observing team, and to the very poor observing conditions
during some winter nights.

To estimate black hole masses from reverberation data, the second
moment of the \hbeta\ line in the rms spectrum is most often used as
the measure of line width \citep{peterson2004}.  We find
$\sigma$(\hbeta$_\mathrm{rms}$) $ = 1740\pm101$ \kms, where the
measurement uncertainty is determined through a bootstrap resampling
procedure \citep{bentz2009a}, and the line width has been corrected
for instrumental broadening of $\sigma_\mathrm{inst}\approx133$ \kms\
following \citet{barth2011}.

\section{Reverberation Lag Measurements}
\label{reverbsection}

\begin{figure}[t!]
\plotone{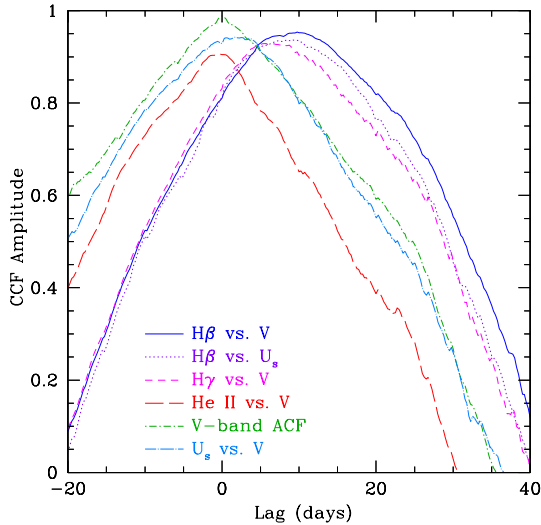}
\caption{Cross-correlation functions for \hbeta, \hgamma, and \heii\
  against the AGN continuum flux, the autocorrelation function of the
  $V$-band continuum, and the cross-correlation of the continuum bands
  $U_s$ vs.\ $V$.
\label{figccf}}
\end{figure}

In order to measure the cross-correlation function (CCF) for unevenly
sampled time series, we employ the interpolation cross-correlation
function methodology and Monte Carlo error analysis techniques
described by \citet{gaskellpeterson1987}, \citet{whitepeterson1994},
and \citet{peterson2004}; these methods have been employed in the
majority of recent reverberation-mapping work
\citep[e.g.,][]{bentz2009a, denney2010}.  We measured the
cross-correlations of \hbeta, \hgamma, and \heii\ against the $V$
light curve, and also of \hbeta\ against the $U_s$ light curve.  The
CCFs were computed from $-20$ to $+40$~days in increments of 0.25
days.  Table \ref{xcortable} lists two measures of the lag: \tpeak,
which is the lag at the peak of the CCF, and \tcen, the centroid of
the CCF for all points above 80\% of the peak value
\citep{peterson2004}.  Figure \ref{figccf} illustrates the CCF
measurements.

The \hbeta\ and \hgamma\ lines have similar lag times of $\tcen =
10.64$ and 8.43~days, respectively, but for \heii\ we find that both
\tcen\ and \tpeak\ are consistent with zero lag, indicating a very
compact size for the inner, high-ionization portion of the BLR.  The
faster response time for \heii\ is apparent in the light curves,
particularly in the steep drop beginning in mid-April (HJD $\approx
2455670$).  The Balmer lines, in contrast, show a much more gradual
and delayed decline in response to the falling continuum flux.  The
flattened peak and asymmetry of the \hbeta\ CCF in comparison with the
continuum autocorrelation function (ACF) suggests that the Balmer-line
emitting zone of the BLR covers a fairly large radial extent, and this
interpretation is supported by the velocity-resolved measurements
described below.  We also measured the CCF between the $U_s$ and $V$
bands in order to search for evidence of reverberation due to Balmer
continuum emission from the BLR \citep{maoz1993,koristagoad2001}, but
no significant lag was found.

\begin{figure}[t!]
\plotone{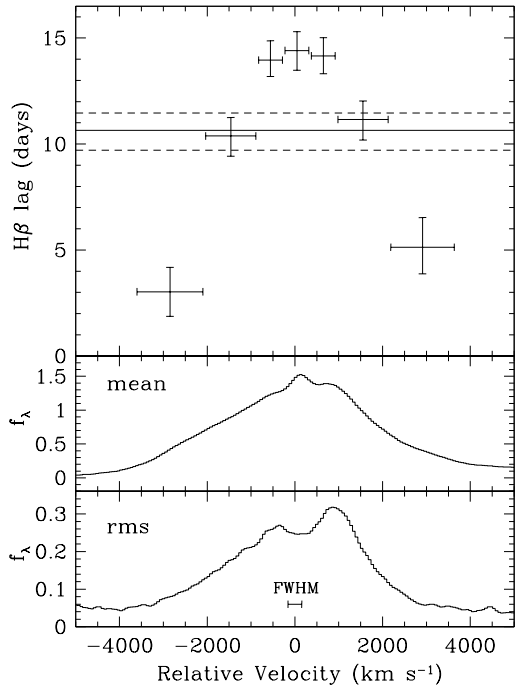}
\caption{Velocity-resolved reverberation in the \hbeta\ line.  The
  upper panel shows the mean lag measured for each velocity segment of
  the broad \hbeta\ line, with the horizontal error bar representing
  the width of the velocity segment.  The overall lag for \hbeta\ and
  its uncertainty range are shown by solid and dashed lines.  The
  lower panels show the mean and rms continuum-subtracted spectra, and
  the error bar indicates the FWHM due to instrumental broadening.
\label{figvdelay}}
\end{figure}

Our Mrk 50 dataset, which benefits from high-amplitude variability and
high-cadence sampling, presents an excellent case study for
velocity-resolved variability. Light curves were measured for seven
separate velocity segments across the width of the \hbeta\ line, and
each segment light curve was cross-correlated against the $V$ light
curve.  Figure \ref{figvdelay} illustrates the lag (\tcen) as a
function of velocity across the \hbeta\ line, revealing a roughly
symmetric trend of shorter lags in the line wings and longer lags in
the core, with a $\sim10$ day difference in response time between the
core and wings.  This pattern, which resembles the symmetric \hbeta\
lag response seen in some other AGNs including Mrk 110
\citep{kollatschny2002} and NGC 5548 \citep{denney2010}, is
qualitatively consistent with predictions for BLR clouds in circular
orbits in the Keplerian potential of the black hole
\citep[e.g.,][]{welsh1991}, with the highest line-of-sight velocities
originating from gas located close to the black hole.  In contrast, a
BLR dominated by either radial infall or outflow would result in an
asymmetric velocity-lag map, with inflow producing longer lags on the
blueshifted side of the line, and outflow producing longer lags on the
red side.  Such signatures of radial motion have been seen in a few
objects \citep{bentz2009a, denney2010}, but the current sample of AGNs
with velocity-resolved data of this quality is still too small to
examine the distribution of different BLR kinematic states.

\begin{deluxetable}{lcc}[t!]
\tablecaption{Cross-Correlation Lag Results}
\tablehead{
  \colhead{Measurement} &
  \colhead{~~~~~~~~~~\tcen\ (days)~~~~~~~~~~} &
  \colhead{\tpeak\ (days)}
}
\startdata
\hbeta\ vs.\ $V$  & $10.64_{-0.93}^{+0.82}$ & $~9.75_{-1.00}^{+0.50}$ \\
\hgamma\ vs.\ $V$ & $~8.43_{-1.28}^{+1.30}$ & $~7.00_{-1.50}^{+1.75}$  \\
\heii\ vs.\ $V$   & $-0.97_{-1.07}^{+1.18}$ & $-0.25_{-1.25}^{+0.75}$  \\
\hbeta\ vs. $U_s$ & $~9.58_{-0.90}^{+1.05}$ & $~8.75_{-1.25}^{+1.00}$ \\
$U_s$ vs. $V$     & $~0.60_{-1.19}^{+1.26}$ & $~1.25_{-1.00}^{+1.50}$ 
\enddata
\tablecomments{All lags are given in the observed frame.}
\label{xcortable}
\end{deluxetable}

\section{Black Hole Mass Estimate}

Following \citet{peterson2004}, we compute the \hbeta\ ``virial
product'' [defined as $\mathrm{VP} = \rblr (\Delta V)^2 / G$] by using
$\Delta V = \sigma$(\hbeta$_\mathrm{rms}$) and $\rblr = c\tcen$, where
\tcen\ has been corrected to the AGN rest frame.  For Mrk 50,
$\rblr=10.40_{-0.91}^{+0.80}$ lt-days, and $\mathrm{VP} =
(6.2\pm0.9)\times10^6$ \msun.

While the virial product is a straightforward combination of measured
quantities, the relationship between virial product and black hole
mass is more indirect and uncertain.  Most recent reverberation work
has adopted a normalization of the virial mass scale based on the
assumption that AGNs as a whole fall on the same \msigma\ relation as
nearby inactive galaxies \citep{onken2004, woo2010}.  Using the
\msigma\ relation derived by \citet{gultekin2009} as the local
reference, this implies a mean value of $\log f =
0.72^{+0.09}_{-0.10}$, and the reverberation masses determined in this
way show an intrinsic scatter of 0.44 dex about the \msigma\ relation
\citep{woo2010}. Adopting the Woo \etal\ value of $f=5.25$ implies
that $\mbh = (3.2\pm0.5)\times10^7$ \msun\ for Mrk 50. For consistency
with recent work \citep{peterson2004, bentz2009a, denney2010}, the
quoted uncertainty includes only the statistical error on the virial
product, but not the uncertainty resulting from the choice of a
specific $f$ value; the true uncertainty in \mbh\ is dominated by the
uncertainty in $f$.  Masses derived in this way depend on the
assumption of virial motion, the assumption that AGNs should fall on
the same \msigma\ relation as quiescent galaxies, and the adoption of
a specific \msigma\ relation as the local baseline.  Differing
assessments of the form of the local \msigma\ relation, particularly
at low masses, can potentially affect the normalization of the AGN
virial mass scale at the factor of $\sim2$ level \citep{greene2010,
graham2011}.  Additionally, there is ongoing debate over the possible
effect of radiation pressure on BLR clouds and its impact on the
inferred masses \citep{marconi2008, netzer2010}. Thus, while the mean
value of $f$ can currently be determined to $\sim20\%$ precision
within the context of a specific set of assumptions about the \msigma\
relation \citep{woo2010, graham2011}, the actual uncertainty in the
overall AGN mass scale remains significantly larger and is difficult
to quantify, and the observed 0.44 dex scatter in the AGN \msigma\
relation must reflect, at least in part, the dispersion of the true
$f$ values for individual AGNs.  Resolving these issues will require
enlarging the number of targets having high-quality
reverberation-mapping data, and the application of new approaches to
extract dynamical information from the observations
\citep[e.g.,][]{pancoast2011a, brewer2011}.

Mrk 50 has an early-type morphology, and \citet{malkan1998} classify
it as an S0 galaxy based on \emph{Hubble Space Telescope} imaging. The
only published measurement of the stellar velocity dispersion of Mrk
50 is $\sigmastar=78\pm15$ \kms\ based on a Sloan Digital Sky Survey
spectrum \citep{greene2006}.  However, a new measurement from Keck
LRIS data gives $109\pm14$ \kms\ (Harris \etal, in preparation), and
we consider this to be more reliable than the SDSS measurement.  The
scaling relations calibrated by \citet{gultekin2009} then imply an
expected $\mbh \approx (0.6-1.7)\times10^7$ \msun\ from the general
\msigma\ relation including ellipticals and spirals, or
$(0.9-2.5)\times10^7$ \msun\ based on the \msigma\ relation fitted to
ellipticals only.  Our reverberation-based mass is slightly higher
than these values, but Mrk 50 still lies well within the $\sim0.4$ dex
scatter of the AGN \msigma\ relation \citep{woo2010}.

\section{Conclusions and Future Work}

From our Spring 2011 monitoring campaign, we have obtained very
high-quality reverberation mapping of Mrk 50.  This is one of just a
few nearby AGNs in which strong velocity-resolved lag signatures have
been detected, and it is now among the most promising targets for
detailed studies of BLR geometry and kinematics.
 
Our long observing campaign produced a large and very rich dataset,
and this paper presents just one set of early results from this
program.  Future papers will include detailed descriptions of the
data-analysis procedures and results for the entire observed sample.
A major goal of our project is to exploit the potential of
velocity-resolved reverberation mapping to elucidate the structure of
the BLR and to derive black hole masses directly, and an upcoming
paper \citep{pancoast2011b} will describe new modeling of our Mrk 50
data. The black hole mass determined from dynamical modeling is
consistent with the simple virial estimate presented here, and Mrk 50
is now the second object \citep[after Arp 151;][]{brewer2011} to show
agreement between the two approaches.

\acknowledgments

We are extremely grateful to the Lick Observatory staff for their
outstanding support during our observing run.  The Lick AGN Monitoring
Project 2011 is supported by NSF grants AST-1107812, 1107865, 1108665,
and 1108835.  T.T. acknowledges a Packard Research Fellowship. The
West Mountain Observatory receives support from NSF grant
AST-0618209. We thank Brad Peterson for a helpful referee report.

\emph{Facilities:} \facility{Shane (Kast)}, \facility{KAIT},
\facility{BYU:0.9m}, \facility{PO:1.5m}, \facility{LCOGT (FTS)},
\facility{Super-LOTIS}

\end{document}